\documentstyle[tighten,preprint,aps,eqsecnum,amssymb,graphicx]{revtex}
\begin{document}
\draft
\title{Hadron formation in high energy photonuclear reactions}

\author{T. Falter and U. Mosel}

\address{Institut fuer Theoretische Physik\\ Universitaet
Giessen\\ D-35392 Giessen, Germany}

\date{March 20, 2002}

\maketitle

\begin{abstract}
We present a new method to account for coherence length effects in a 
semi-classical transport model. This allows us to describe photo- and 
electroproduction at large nuclei ($A>12$) and high energies using a realistic
coupled channel description of the final state interactions that goes 
beyond simple Glauber theory. We show that the purely absorptive treatment of 
the final state interactions can lead to wrong estimates of color transparency 
and formation time effects in particle production. As an example, we discuss 
exclusive $\rho^0$ photoproduction on Pb at a photon energy of 7~GeV as well as
$K^+$ production in the photon energy range 1-7~GeV.
\end{abstract}


\section{Introduction} \label{sec:intro}
In a high energy collision between two hadrons or a photon and a hadron it 
takes a finite amount of time for the reaction products to evolve to physical 
particles. During the collision process some momentum transfer between the 
hadrons or some hard scattering between two of the hadrons' constituents 
leads to the excitation of hadronic strings. The time that is needed for the 
creation and fission of these strings as well as for the hadronization of the
string fragments cannot be calculated within perturbative QCD because the 
hadronization process involves small momentum transfers of typically only a few
hundred MeV. One can perform an estimate of the formation time $\tau_f$ in the 
rest frame of the hadron. It should be of the order of the time that the 
quark-antiquark (quark-diquark) pair needs to reach a separation, that is of 
the size of the produced hadron ($r_h\approx 0.6-0.8$~fm):
\begin{equation}  
  \tau_f\gtrsim\frac{r_h}{c}.\\
\end{equation}

During their evolution to physical hadrons the reaction products will react 
with reduced cross sections. This is motivated by means of color transparency: 
the strings and the substrings created during the fragmentation are in a color 
singlet state and therefore react with a cross section that increases with 
their transverse size. 
As a consequence the produced hadrons travel inside the nuclear medium with a
reduced scattering probability during their formation time. Hence the formation
time plays an important role in the dynamics of nuclear reactions, e.g. heavy 
ion collisions, proton and pion induced reactions as well as photon and 
electron induced reactions on nuclei.\\

The latter two are of special interest because they are less complex than 
heavy ion collisions and, in contrast to hadron induced reactions, the primary 
reaction does in general not only take place at the surface of the nucleus but 
also at larger densities. Experiments at TJNAF and DESY, for example, deal with
exclusive and semi-inclusive meson photo- and electroproduction at high 
energies. Large photon energies $E_\gamma$ are of special interest because of 
time dilatation the formation length $l_f$ in the rest frame of the nucleus can
exceed nuclear dimensions:
\begin{equation}
  l_f=v_h\cdot\gamma\cdot\tau_f=\frac{p_h}{m_h}\cdot\tau_f.
\end{equation}
If one chooses the formation time to be $\tau_f=0.8$~fm/c, the formation length
in the rest frame of the nucleus will be about 30~fm for a 5~GeV pion and about
7~fm for a 5~GeV kaon or a 7~GeV $\rho$ meson. These lengths have to be
compared with the typical size of nuclear radii, e.g. 2.7~fm for $^{12}$C and 
7.1~fm for $^{208}$Pb. Because it suppresses the final state interactions, the 
formation time has a big effect on photonuclear production cross sections at 
high energies. Turning this argument around, exclusive and semi-inclusive 
photoproduction of mesons on nuclei~\cite{HERMES01} offer a possibility to 
study these formation times if the FSI are well under control.\\
 
Usually, the FSI are modeled within Glauber theory~\cite{Gla59Gla70,Yen71} and 
treated purely absorptive. 
A more realistic description of the FSI has to take also regeneration of the 
mesons studied into account. For this we use a coupled channel semi-classical 
transport model based on the Boltzmann-Uehling-Uhlenbeck (BUU) equation.
Originally developed to describe heavy ion collisions~\cite{SIS} at SIS 
energies it has been extended in later works to investigate also inclusive 
particle production in heavy ion collisions up to 200~AGeV and 
$\pi$~\cite{Eff99a} and $p$ induced as well as photon and electron induced 
reactions in the resonance region~\cite{res-reg}. Inclusive photoproduction of 
mesons at energies between 1 and 7~GeV has been investigated in~\cite{Eff00}. 
An attractive feature of this model is its capability to describe a large 
variety of very different reaction types in a consistent way.\\

Since photon induced reactions are known to be shadowed 
($\sigma_{\gamma A}<A\sigma_{\gamma N}$) above photon energies of approximately
1~GeV~\cite{Bia96Muc99,Fal00,Fal01}, one needs a way to
account for this shadowing effect in photoproduction. This is straightforward 
within Glauber theory, but it is not clear how to account for this initial 
state coherence length effect in a semi-classical transport model for the FSI.
A first attempt at combining the quantum mechanical coherence in the 
entrance channel with our incoherent treatment of the FSI has been made 
in~\cite{Eff00}. In Sec.~\ref{sec:model} we present a new, improved way to 
implement shadowing in our model. In Sec.~\ref{sec:results} the results for 
exclusive $\rho^0$ photoproduction on $^{208}$Pb at 7~GeV are compared with 
the predictions from simple Glauber theory. In Sec.~\ref{sec:results} we also 
demonstrate the importance of the coupled channel treatment of the FSI by 
looking at the nuclear $K^+$ photoproduction cross section. We summarize our 
results in Sec.~\ref{sec:summary}.\\

\section{Model}\label{sec:model}
In this section we describe how we account for shadowing in our model and 
only sketch the principles of the transport model itself. For a detailed 
description of the transport model used we refer to~\cite{Eff99b}.
In our model the incoherent reaction of a 
high energy photon with a nucleus takes place in two steps. In the first step 
the photon reacts with one nucleon inside the nucleus (impulse approximation) 
and produces some final state $X$. In this process nuclear effects like Fermi 
motion, binding energies and Pauli blocking of the final state nucleons are 
taken into account. In the second step the final state $X$ is propagated within
the transport model. Except for the elastic vector meson and exclusive 
strangeness production (see~\cite{Eff00}) we use the Lund string model 
FRITIOF~\cite{And93} to describe high energy photoproduction on the 
nucleon. The same string model is also used to deal with high energy particle 
collisions in the FSI. Since FRITIOF does not accept photons as incoming 
particles, $\rho$ dominance was used in~\cite{Eff00} and the photon was passed 
as a massless $\rho^0$. This led to an excellent description of charged 
particle multiplicities in $\gamma p$ collisions.\\

Here we generalize to vector meson dominance (VMD)~\cite{Bau78} by writing the
incoming photon state as
\begin{equation}
\label{eq:vmd}
  |\gamma\rangle=\left(1-\sum_{V=\rho,\omega,\phi}\frac{e^2}{2g_V^2}\right)|
\gamma_0\rangle+\sum_{V=\rho,\omega,\phi}\frac{e}{g_V}|V\rangle,
\end{equation}
and pass the photon as a massless $\rho^0$, $\omega$ or $\phi$ with a 
probability corresponding to the strength of the vector meson coupling to the 
photon times its nucleonic cross section $\sigma_{VN}$
\begin{equation}
  P(V)=\frac{\left(\frac{e}{g_V}\right)^2\sigma_{VN}}{\sigma_{\gamma N}}.
\end{equation}
These probabilities follow from using (\ref{eq:vmd}) in the optical theorem:
\begin{equation}
  \label{eq:opti}
  \sigma_{\gamma N}=\left(1-\sum_{V=\rho,\omega,\phi}\frac{e^2}{2g_V^2}\right)^2\sigma_{\gamma_0 N}+\sum_{V=\rho,\omega,\phi}\left(\frac{e}{g_V}\right)^2\sigma_{VN}.
\end{equation} 
As can be seen from (\ref{eq:vmd}) and (\ref{eq:opti}) there is also a finite 
probability 
\begin{equation}
  P(\gamma_0)=1-\sum_{V=\rho,\omega,\phi}P(V)
\end{equation}
that the 'bare' photon $\gamma_0$ has to be passed to FRITIOF; the component
$\gamma_0$ does not get shadowed in the nucleus. In a generalized VMD model for
example the $\gamma_0$ includes 
contributions from heavy intermediate hadronic states which, as we show later, 
are not shadowed because of the large momentum transfer $q_V$ that is needed 
to put them on their mass shell. Since FRITIOF does not accept a 'bare' photon
as input we replace it again by a vector meson 
$V$(=$\rho^0$, $\omega$ or $\phi$), with the probability
\begin{equation}
\label{eq:baregam}
  P_{\gamma_0}(V)=\frac{\left(\frac{e}{g_V}\right)^2\sigma_{VN}}{\sum_{V'=\rho,\omega,\phi}\left(\frac{e}{g_{V'}}\right)^2\sigma_{V'N}}.\\
\end{equation}

The particle production in FRITIOF can be decomposed into two parts. First 
there is a momentum transfer taking place between the two incoming
hadrons leaving two excited strings with the quantum numbers of the initial 
hadrons. After that the two strings fragment into the observed particles. As a 
formation time we use 0.8~fm/c in the rest frame of each hadron; during this 
time the hadrons do not interact with the rest of the system.\\

\begin{figure}
  \begin{center}
    \includegraphics[width=8cm]{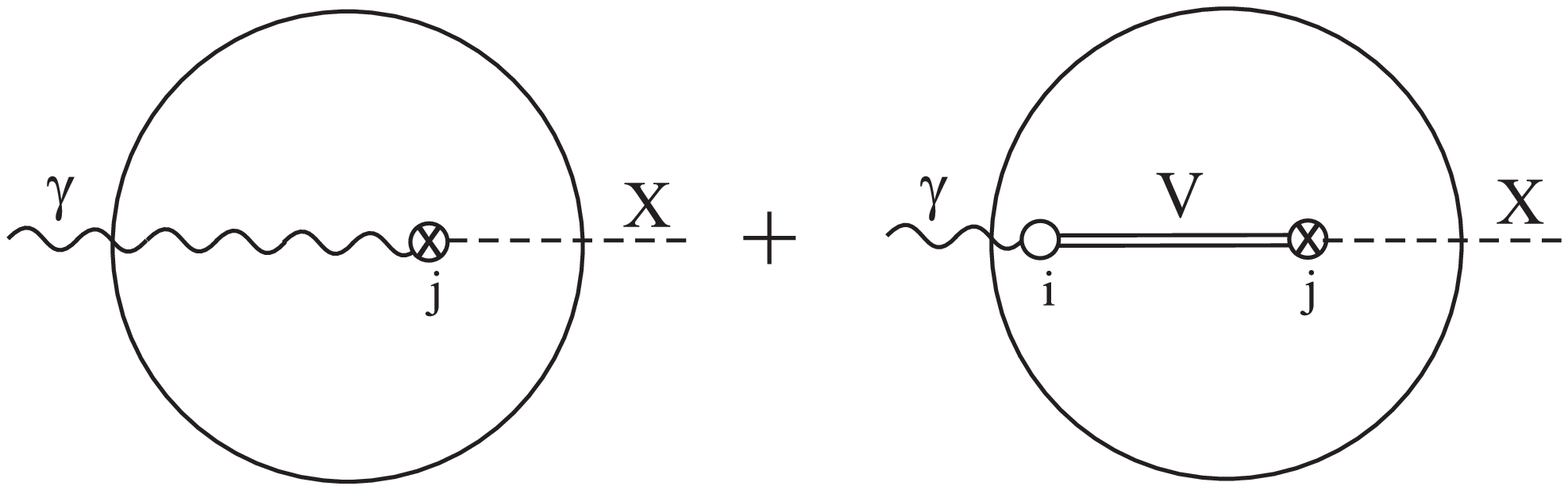}
  \end{center}
  \caption{The two amplitudes of order $\alpha_{em}$ that contribute to 
incoherent meson photoproduction in simple Glauber theory. The left amplitude
alone would lead to an unshadowed cross section. Its interference with the
right amplitude gives rise to shadowing.}
  \label{fig:Glauber}
\end{figure}

Up to now we did not take shadowing into account. In Glauber theory this 
corresponds to using only the left amplitude in Fig.~\ref{fig:Glauber}, where a
photon directly produces some hadron $X$ at nucleon $j$.
The left amplitude alone leads to the unshadowed incoherent 
photoproduction cross section
\begin{equation}
\label{eq:unshad}
  \sigma_{\gamma A\rightarrow XA^*}^{\text{unshadowed}}=\sigma_{\gamma N\rightarrow XN}\int d^2b\int dz_j\tilde{n}(\vec{b},z_j)e^{-\sigma_{XN}\int_{z_j}^\infty dz n(\vec b,z)}
\end{equation}
where $n(\vec r)$ denotes the nucleon number density, $\tilde{n}(\vec r)$ 
the number density of nucleons with the correct charge to produce the hadron 
$X$ and $\sigma_{XN}$ is the total $XN$ cross section. The exponential damping
factor in (\ref{eq:unshad}) describes the absorption of the particle $X$ on 
its way out of the nucleus.\\

In Glauber theory shadowing arises from the 
interference of the left amplitude in Fig.~\ref{fig:Glauber} with the second 
amplitude of order $\alpha_{em}$ which is shown on the right hand side. In this
process the photon produces a vector meson $V$ on nucleon $i$ without 
excitation of the nucleus. 
This vector meson then scatters at fixed impact parameter $\vec b$ 
(eikonal approximation) through the nucleus to nucleon $j$ and produces the 
final state meson $X$, leaving the nucleus in the same excited state as in the 
direct process. Off-diagonal scattering, where a vector meson $V$ scatters into
a different vector meson $V'$, is usually neglected. In view of 
the upcoming discussion of exclusive $\rho^0$ photoproduction on Pb we state 
here the Glauber result~\cite{Yen71} for the incoherent vector meson production
cross section:
\begin{eqnarray}
  \label{eq:Glauber}
  \sigma_{\gamma A\rightarrow VA^*}&=&\sigma_{\gamma N\rightarrow VN}\int d^2b\int_{-\infty}^{\infty}dz_j n(\vec b,z_j)e^{-\sigma_{VN}\int_{z_j}^\infty dz n(\vec b,z)}\nonumber\\
& &\times\left|1-\int_{-\infty}^{z_j}dz_i n(\vec b,z_i)\frac{\sigma_{VN}}{2}(1-i\alpha_V)e^{iq_Vz_i}\exp\left[-\frac{1}{2}\sigma_{VN}(1-i\alpha_V)\int_{z_i}^{z_j}dzn(\vec b,z)\right]\right|^2.
\end{eqnarray}
Again $n(\vec r)$ denotes the nucleon number density, $\sigma_{VN}$ the total 
$VN$ cross section and $\alpha_V$ the ratio of real to imaginary part of the 
$VN$ forward scattering amplitude. The momentum transfer 
\begin{equation}
  q_V\approx\frac{m_V^2}{2E_\gamma}
\end{equation}
arises from putting the vector meson on its mass shell. A large momentum 
transfer $q_V$ is suppressed by the elastic nuclear formfactor and leads to 
less shadowing due to the oscillating factor $\exp[i q_Vz_i]$ 
in the integrand of Eq.~(\ref{eq:Glauber}). Note that $q_V$ is just the inverse
of the coherence length, i.e. the distance, given by the uncertainty 
principle, which the photon can travel as a $V$ fluctuation. For the
quantities in (\ref{eq:Glauber}) we use the parameterizations of Model~I of 
Ref.~\cite{Bau78} with which one obtains a very good description of the 
shadowing effect in nuclear photoabsorption down to the onset 
region~\cite{Fal00,Fal01}.\\

\begin{figure}
  \begin{center}
    \includegraphics[width=10cm]{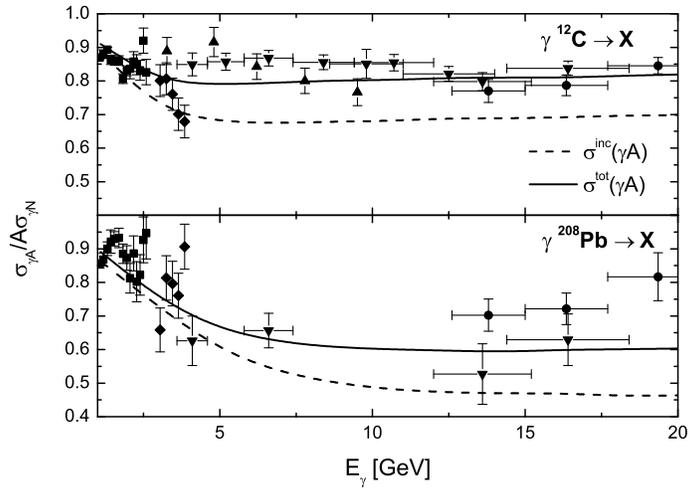}
  \end{center}
  \caption{The nuclear photoabsorption cross section divided by 
$A\sigma_{\gamma N}$ plotted versus the photon energy $E_\gamma$. 
The solid line represents the result of Reference~\protect\cite{Fal00} 
for the total photon nucleus cross section. The dashed line shows the 
contribution from incoherent reactions to $\sigma_{\gamma A}^{tot}$ and is 
calculated using Eq.~(\ref{eq:intinc}). More than 90\% of the difference 
between $\sigma_{\gamma A}^{inc}$ and $\sigma_{\gamma A}^{tot}$ is due to 
coherent $\rho^0$ photoproduction.}
  \label{fig:abscs}
\end{figure}

One clearly sees from (\ref{eq:Glauber}) how the FSI separate from the 
'initial state interactions' of the photon. We now replace the exponential 
damping factor 
$\exp [-\sigma_{VN}\int_{z_j}^\infty dz n(\vec b,z)]$, which corresponds to 
purely absorptive FSI, by a transport model. This allows us to incorporate a 
wider class of FSI. In addition we want to include events where the final 
vector meson $V$ is not produced in the primary reaction but via sidefeeding. 
This leads to less reduction of the nuclear production cross section even with 
the same $\sigma_{VN}$. We thus also need to know how all the other possible 
primary reactions are shadowed. 
We, therefore, start from (\ref{eq:vmd}) and use Glauber theory to calculate 
how the single $V$ components of the photon change due to multiple scattering 
on the way to nucleon $j$ where the state $X$ is 
produced~\cite{Yen71}:
\begin{equation}
  \label{eq:gamma}
  |\gamma(\vec r_j)\rangle=\left(1-\sum_{V=\rho,\omega,\phi}\frac{e^2}{2g_V^2}\right)|\gamma_0\rangle+\sum_{V=\rho,\omega,\phi}\frac{e}{g_V}\left(1-\overline{\Gamma_V}(\vec r_j)\right)|V\rangle.
\end{equation}
Here the (photon energy dependent) nuclear profile functions for the different 
vector meson components $V$ are given as
\begin{equation}
  \overline{\Gamma_V}(\vec b,z_j)=\int_{-\infty}^{z_j}dz_in(\vec b,z_i)\frac{\sigma_{VN}}{2}(1-i\alpha_V)e^{iq_V(z_i-z_j)}\exp\left[-\frac{1}{2}\sigma_{VN}(1-i\alpha_V)\int_{z_i}^{z_j}dzn(\vec b,z)\right].
\end{equation}
Note that the $\gamma_0$ component is, by definition, not modified due to the 
presence of the nuclear medium. The cross section for the photon to react with 
nucleon $j$ at position $\vec r_j$ inside the nucleus can be deduced via 
(\ref{eq:gamma}) from the optical theorem:
\begin{equation}
\label{eq:inccs}
  \sigma_{\gamma N}(\vec r_j)=\left(1-\sum_{V=\rho,\omega,\phi}\frac{e^2}{2g_V^2}\right)^2\sigma_{\gamma_0 N}+\sum_{V=\rho,\omega,\phi}\left(\frac{e}{g_V}\right)^2\left|1-\overline{\Gamma_V}(\vec r_j)\right|^2\sigma_{VN}.
\end{equation} 
Like for the photon in vacuum each term gives the relative weight for the 
corresponding photon component to be passed to FRITIOF. When integrated over 
the whole nucleus one gets from Eq.~(\ref{eq:inccs}) the total incoherent 
photonuclear cross section
\begin{equation}
  \label{eq:intinc}
  \sigma_{\gamma A}^{inc}=\int d^3r_jn(\vec r_j)\sigma_{\gamma N}(\vec r_j)
\end{equation}
which is shown in Fig.~\ref{fig:abscs} by the dashed line together with the 
total nuclear photoabsorption cross section as derived from the optical theorem
in~\cite{Fal00}:
\begin{eqnarray}
\label{eq:cs}
    \sigma^{tot}_{\gamma A}&=&A\sigma_{\gamma N}-\sum_{V=\rho,\omega,\phi}\frac{k_V}{2k}\left(\frac{e}{g_V}\right)^2\textrm{Re}\biggl\{\sigma_{VN}^2(1-\alpha_V)^2\int d^2b\int_{-\infty}^{\infty}dz_i\int_{z_i}^{\infty}dz_jn(\vec b,z_i)n(\vec b,z_j)\nonumber\\
& &\quad\times
e^{iq_V(z_i-z_j)}\exp\left[-\frac{1}{2}\sigma_{VN}(1-i\alpha_V)\int_{z_i}^{z_j}dz'n(\vec
b,z')\right]\biggr\}.
\end{eqnarray}
Here $k$ and $k_V$ denote the momentum of the photon and the vector meson 
respectively and we again use the parameterizations of Model~I of 
Ref.~\cite{Bau78}. More than 90\% of the difference between those two cross 
sections stems from coherent $\rho^0$ photoproduction.\\ 

\begin{figure}
  \begin{center}
     \includegraphics[width=16cm]{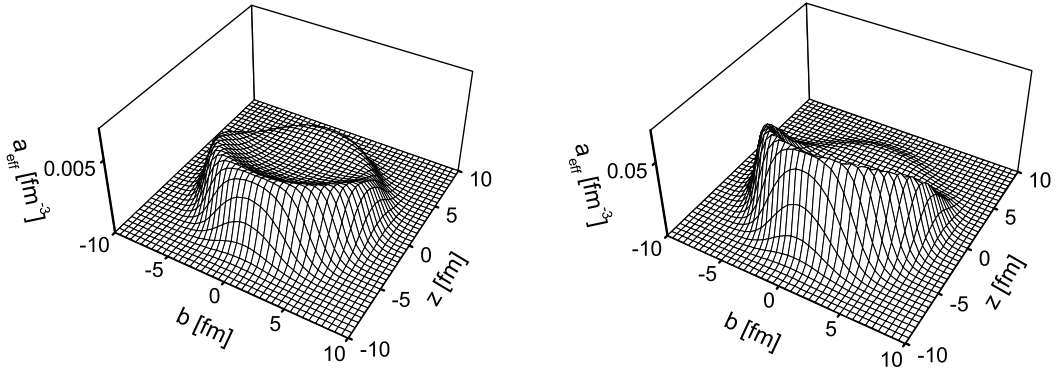}
  \end{center}
  \vspace{-2cm}
   \caption{The number density of nucleons that react with the $\phi$
component ({\it left side}) and $\rho^0$ component ({\it right side}) of a
20~GeV photon for $^{208}$Pb calculated using Eq.~(\ref{eq:aeffdens})
. In both 
cases the nucleons on the front side of the nucleus shadow the downstream 
nucleons. This effect is stronger for the $\rho^0$ component because of its 
larger nucleonic cross section.}
  \label{fig:aeff}
\end{figure}

In Fig.~\ref{fig:aeff} we show how strongly the $\rho^0$ and the $\phi$ 
components of a real 20~GeV photon are separately shadowed in Pb. We plot the 
number density of the nucleons reacting with the $V$ components of the photon
\begin{equation}
  \label{eq:aeffdens}
  a_{eff}^V(\vec r_j)=n(\vec r_j)\frac{1}{\sigma_{\gamma N}}\left(\frac{e}{g_V}\right)^2\left|1-\overline{\Gamma_V}(\vec r_j)\right|^2\sigma_{VN}
\end{equation}
as a function of $\vec r_j$. One clearly sees that due to its smaller 
nucleonic cross section the $\phi$ component is less shadowed than the
$\rho^0$ component at the backside of the nucleus. As a consequence strangeness
production (e.g. $K$ photoproduction), where the primary reaction is preferably
triggered by the $\phi$ component of the photon, is less shadowed than, e. g., 
$\pi$ photoproduction. This dependence of the strength of 
shadowing on the reaction type is new compared to the treatment of shadowing 
in~\cite{Eff00} and can also be seen directly from the second amplitude in 
Fig.~\ref{fig:Glauber} because of the occurrence of the scattering process 
$VN\rightarrow XN$ at nucleon $j$.\\

As already mentioned above the purely absorptive FSI of the Glauber model
are very different from the coupled channel description of a transport model.
The transport model we use is based on the BUU equation that describes the time
evolution of the phase space density $f_i(\vec r,\vec p,t)$ of particles of 
type $i$ that can interact via binary reactions~\cite{Eff99b}. In our case 
these particles are the nucleons of the target nucleus as well as the baryonic 
resonances and mesons ($\pi$, $\eta$, $\rho$, $K$, ...) that can either be 
produced in the primary $\gamma N$ reaction or during the FSI. For particles of
type $i$ the BUU equation looks as follows:
\begin{equation}
  \left(\frac{\partial}{\partial t}+\frac{\partial H}{\partial\vec r}\frac{\partial}{\partial \vec r}-\frac{\partial H}{\partial \vec r}\frac{\partial}{\partial \vec p}\right)f_i(\vec r,\vec p,t)=I_{coll}[f_1,...f_i,...,f_M].
\end{equation}
In the case of baryons the Hamilton function $H$ includes a mean field 
potential which in our model depends on the particle position and momentum. 
The collision integral on the right hand side accounts for the creation and
annihilation of particles of type $i$ in a collision as well as elastic 
scattering from one position in phase space into another. For fermions Pauli 
blocking is taken into account in $I_{coll}$ via blocking factors. For each 
particle type $i$ such a BUU equation exists; all are coupled via the mean 
field and the collision integral. This leads to a system of coupled 
differential-integral equations which we solve via a test particle ansatz for 
the phase space density (for details see~\cite{Eff99b}).
Since the collision integral also accounts for particle creation in a collision
the observed outgoing particle $X$ cannot only be produced in the primary 
reaction, but can also be created by sidefeeding in which a particle $Y$ is 
created first which propagates and then, by FSI, produces $X$. In addition the 
state $X$ might get absorbed on its way out of the nucleus but be fed in again 
in a later interaction. Both cases can a priori not be ignored, but are usually
neglected in Glauber models.\\

\section{Results}\label{sec:results}
Exclusive vector meson photo- and electroproduction on nuclei is an ideal tool 
to study effects of the coherence length, formation time and color 
transparency. Exclusive $\rho^0$ electroproduction has been investigated in 
the HERMES experiment~\cite{Ack00} at photon energies between 10~GeV and 20~GeV
and virtuality $Q^2\lesssim$~5~GeV$^2$. 
The calculations for meson production on nuclei are usually done within 
Glauber theory~\cite{Glauber-models}. As already mentioned above the FSI in 
Glauber theory are usually purely absorptive. This means that for the reaction 
$\gamma A\rightarrow\rho^0 A^*$ the primary reaction has to be 
$\gamma N\rightarrow\rho^0 N$. If one treats the FSI via an absorptive optical 
potential one gets an exponential damping 
$\sim\exp [-\sigma_{\rho N}\int_{z_j}^\infty dz n(\vec b,z)]$ of the nuclear
production cross section.\\

\begin{figure}
  \begin{center}
    \includegraphics[width=10cm]{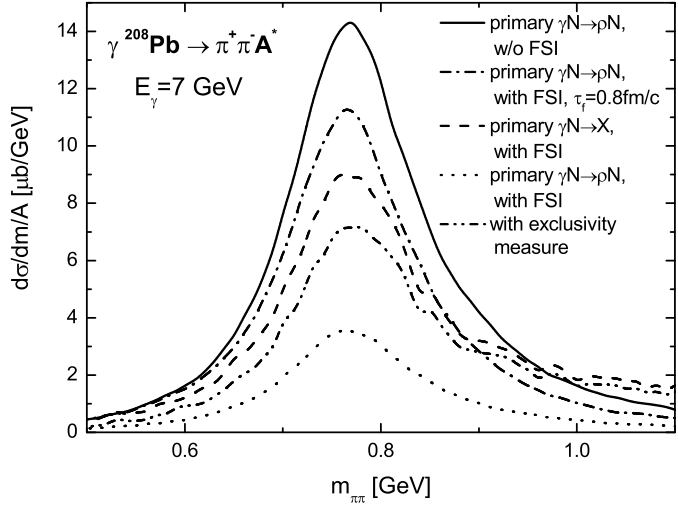}
  \end{center}
  \caption{Mass differential cross section for exclusive $\rho^0$ production
on $^{208}$Pb at $E_\gamma=7$~GeV. The meaning of the different curves is 
explained in detail in the text. All the curves, except the one with the
explicitly given formation time $\tau_f=0.8$~fm/c have been calculated with
$\tau_f=0$. The fluctuations in the dashed and dash-dot-dotted curves are 
statistical only.}
  \label{fig:exclrho}
\end{figure}

We are presently working at incorporating also incoming virtual photons
into the formalism developed in Sec.~\ref{sec:model} and at enlarging the
configuration space for the FSI. Here we discuss, therefore, only results
obtained with real photons and at a lower energy.
In Fig.~\ref{fig:exclrho} we show the results of our model for 
the mass differential cross section of incoherent, exclusive $\rho^0$ 
photoproduction on $^{208}$Pb for $E_\gamma=7$~GeV. In this case 'exclusive' 
means that the final state consists of a $\pi^+\pi^-$ pair and 208 bound 
nucleons. The solid line represents a calculation with the primary reaction 
restricted to $\gamma N\rightarrow\rho^0N$. It already includes the effects
of shadowing, Fermi motion, Pauli blocking and the nucleon potential, but no 
FSI. The dotted line shows the effect of the FSI without a formation time of
the $\rho^0$ in $\gamma N\rightarrow\rho^0N$. The Glauber model yields
quantitatively the same effect of the FSI. This means that in this case (only
$\gamma N\rightarrow\rho^0N$ as primary reaction) FSI 
processes like $\rho^0N\rightarrow\pi N$, $\pi N\rightarrow\rho^0N$, where the
primary $\rho^0$ gets absorbed first and is fed into the outgoing channel by a
later FSI, are negligible. If one assumes a formation time of $\tau_f=0.8$~fm/c
for the $\rho^0$, one gets the result indicated by the dash-dotted line. Due to
the finite formation time there is considerably less absorption and the nuclear
production cross section increases. If the observed spectrum looked like this, 
one would in Glauber theory be lead to the conclusion of a finite $\rho^0$ 
formation time
.\\ 

However, one will get a similar result with $\tau_f=0$ if one allows for other 
primary reaction besides $\gamma N\rightarrow\rho^0N$ and uses a coupled 
channel model. This can be seen by looking at the dashed line in 
Fig.~\ref{fig:exclrho}. We find that about 60\% of the additional $\rho^0$ 
stem from inelastic $\rho^0$ production in the primary reaction, e.g. 
$\gamma N\rightarrow\rho^0\pi N$ where the $\pi$ gets absorbed 
during the FSI. We now apply an exclusivity measure like the one used in the 
HERMES experiment~\cite{Ack00}
\begin{equation}
  \label{eq:exclms}
  -2\text{~GeV}<\Delta E=\frac{p_Y^2-m_N^2}{2m_N}<0.6\text{~GeV},
\end{equation}
where $m_N$ is the nucleon mass and 
\begin{equation}
  \label{eq:py}
  p_Y=p_N+p_{\gamma}-p_{\rho}
\end{equation}
is the 4-momentum of the undetected final state.
Here $p_{\gamma}$ and $p_{\rho}$ denote the 4-momenta of the 
incoming photon and the detected $\pi^+\pi^-$ pair and $p_N$ is the 4-momentum 
of the struck nucleon which, for the calculation of $p_Y$, is 
assumed to be at rest.
Using (\ref{eq:exclms}) leads to a decrease of the cross section 
(dash-dot-dotted line) because 
some of the inelastic primary events are excluded. If the exclusivity measure 
was good enough to single out only the elastic $\gamma N\rightarrow\rho^0N$
reactions from the primary events, the dash-dot-dotted curve would coincide 
with the dotted line and Glauber theory would be applicable. Since this is not 
the case, one still extracts a too large formation time when using Glauber 
theory.\\
 
\begin{figure}
  \begin{center}
    \includegraphics[width=10cm]{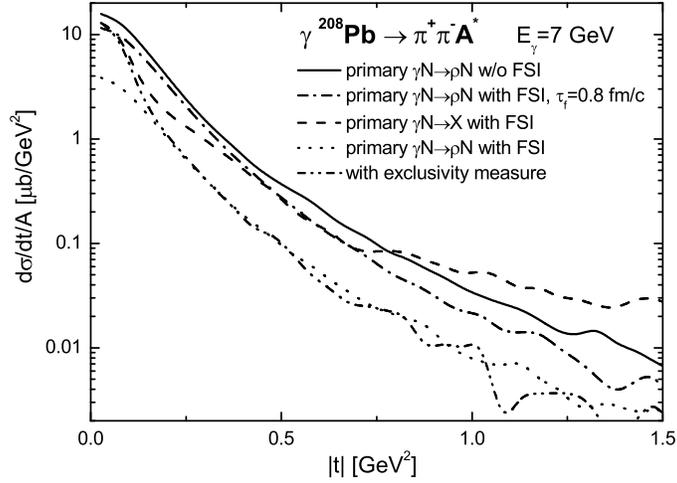}
  \end{center}
  \caption{Calculated $\frac{d\sigma}{dt}$ for exclusive $\rho^0$ production
on $^{208}$Pb at $E_\gamma=7$~GeV. The meaning of the different curves is the 
same as in Fig.~\ref{fig:exclrho}. The structures in the curves at large $|t|$ are
statistical only.}
  \label{fig:exclrho2}
\end{figure}

One therefore needs a further constraint in addition to the exclusivity measure
(\ref{eq:exclms}) which becomes apparent by looking at the differential cross 
section $\frac{d\sigma}{dt}$ in Fig.~\ref{fig:exclrho2}; the meaning of the 
lines is as before. For $|t|>0.1$~GeV$^2$ the full calculation with 
exclusivity measure (dash-dot-dotted line) gives the same result as the one 
with the primary reaction $\gamma N\rightarrow\rho^0N$ and FSI (dotted line). 
In this kinematic regime Glauber theory can therefore be used. 
For $|t|<0.1$~GeV$^2$, however, the exclusivity measure (\ref{eq:exclms}) 
cannot distinguish between elastic $\rho^0$ photoproduction 
($\gamma N\rightarrow\rho^0N$) and other primary reactions, e.g. inelastic 
$\rho^0$ photoproduction 
($\gamma N\rightarrow\rho^0X$, $X\neq N$). At low values of $|t|$ there exist
many states $X$ with invariant masses which are not excluded by the exclusivity
measure (\ref{eq:exclms}).
In addition we find that about 25\% of the finally accepted $\rho^0$ in this
$t$ region are not produced in the primary reaction but stem from sidefeeding
in the FSI. In the HERMES experiment one makes a lower $|t|$ cut to get rid of 
the coherent $\rho^0$ photoproduction contribution. In the case of lead and 
$E_\gamma=7$~GeV the coherent part can be neglected above 
$|t|\approx0.05$~GeV$^2$. For Glauber theory to be reliable one has to 
increase this threshold to approximately $|t|=0.1$~GeV$^2$. Glauber theory can 
thus be trusted only under certain kinematic constraints.\\

\begin{figure}
  \begin{center}
    \includegraphics[width=10cm]{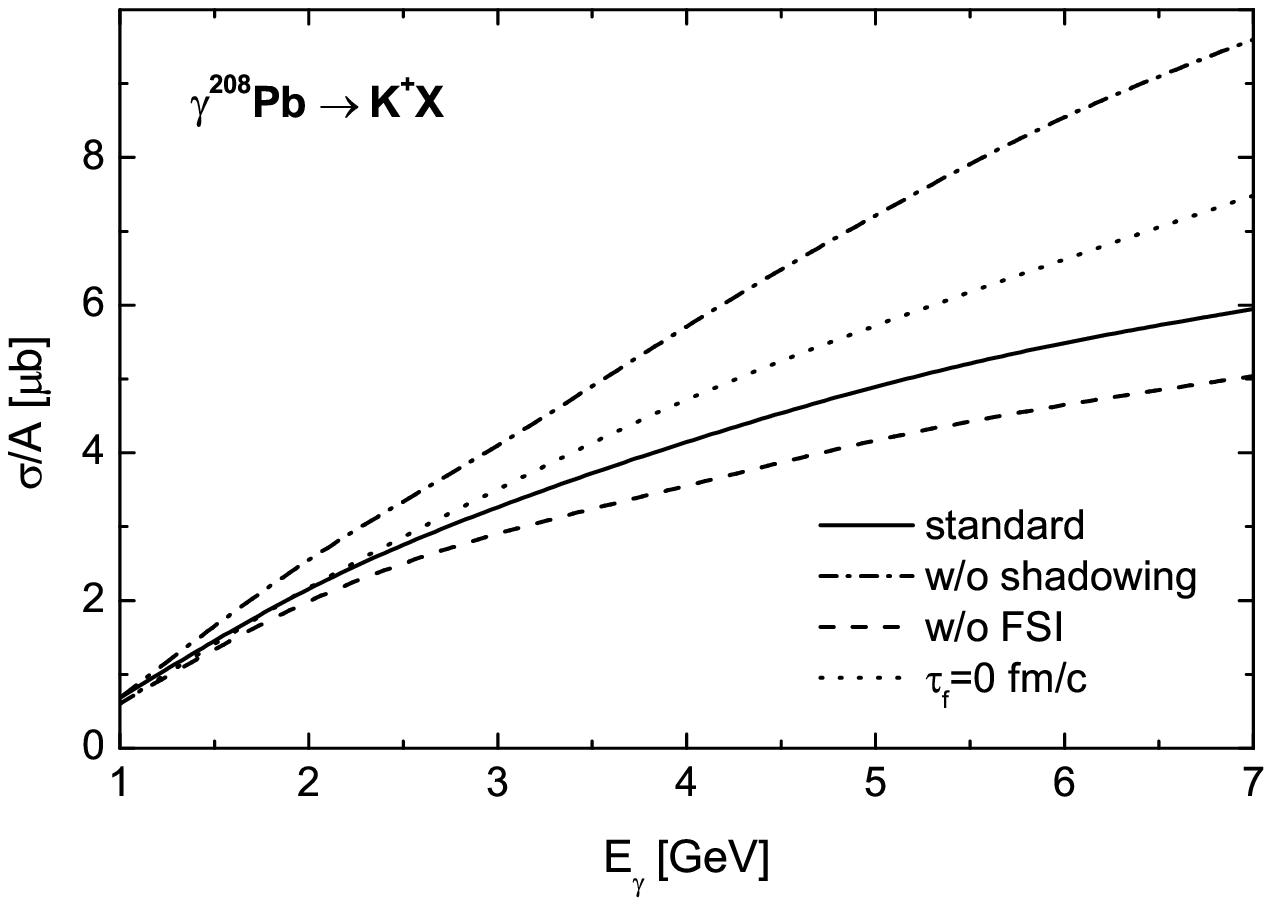}
  \end{center}
  \caption{Photoproduction cross section for $K^+$ on $^{208}$Pb 
plotted as a function of the photon energy. The solid
line represents the full calculation. The dash-dotted line shows the result 
without shadowing of the incoming photon, the dashed line the result without
FSI and the dotted line the calculation without formation time.}
  \label{fig:kaons}
\end{figure}

So far we have discussed a case in which a strongly interacting particle,
the $\rho^0$ meson, is produced. In the following we will now discuss the
special effects that appear in a coupled channel treatment of the FSI when
a weakly interacting particle, such as the $K^+$ meson, is considered.
In Fig.~\ref{fig:kaons} we therefore show the cross section for 
the reaction $\gamma A\rightarrow K^+X$ in the photon energy range 1-7~GeV for 
$^{208}$Pb. This reaction had already been investigated in~\cite{Eff00}; the 
new treatment of shadowing and the initial interactions as outlined in Eqs. 
(\ref{eq:vmd})-(\ref{eq:baregam}) lead to an increase of the 
total yield at 7 GeV by about 20\%. The solid curve in Fig.~\ref{fig:kaons} 
represents the results of the full calculation (including 
shadowing, FSI, $\tau_f=0.8$~fm/c, etc.). By comparison with the calculation 
without shadowing (dash-dotted line) one sees how important shadowing becomes 
with increasing photon energy. At 7~GeV it reduces the nuclear production 
cross section to about 65\%.\\ 

The importance of a coupled channel treatment of the FSI becomes clear 
when comparing the full calculation with the one without FSI (dashed line). 
Since the $\overline{s}$ quark cannot be absorbed in medium the FSI can just 
increase the $K^+$ yield via processes like $\pi N\rightarrow K^+Y$ 
($Y=\Sigma,\Lambda$) for example.
With decreasing formation time the primarily produced pions have a greater 
chance to produce $K^+$ in the FSI. As a consequence of this, a shorter 
formation time will lead to an increase of the nuclear $K^+$ photoproduction 
cross section as can be seen from the dotted line. An enhancement of the $K^+$ 
production cross section due to FSI cannot be explained by purely absorptive 
FSI as in simple Glauber theory where it would necessarily be interpreted
as being due to a longer formation time of the $K^+$.\\

\section{Summary}\label{sec:summary}
High energy photoproduction off nuclei offers a promising 
possibility to study the physics of hadron formation. Necessary for such 
studies is a reliable model of the FSI to extract the formation time from the 
production cross sections. Whereas Glauber models allow for a straightforward
implementation of the nuclear shadowing effect they usually have the 
disadvantage of a purely absorptive treatment of the FSI. As we have shown the
latter can lead to a wrong estimate of the formation time. A more realistic 
treatment of the FSI is possible within a coupled channel transport model. 
We have presented a method to combine such an incoherent treatment of FSI with 
coherence length effects in the entrance channel which can easily be extended 
to higher energies and virtual photons. We have shown that in particular the
production of mesons with long mean free path will be affected by the coupled 
channel effects in the FSI.

\section{Acknowledgments}
This work was supported by DFG.\\


\newpage

\newpage
 
\begin{thebibliography}{99}

\bibitem{HERMES01} A.~Airapetian {\it et al.}, Eur. Phys. J. {\bf C 20}, 479 (2001).

\bibitem{Gla59Gla70} R.~J.~Glauber, in {\it Lectures in Theoretical Physics}, edited by W.E.~Brittin and L.G.~Dunham (Wiley Intersience, New York, 1959), Vol. I, p. 315; R.~J.~Glauber, in {\it High Energy Physics and Nuclear Structure}, edited by S.~Devons (Plenum, New York, 1970), p. 207.

\bibitem{Yen71} D.~R.~Yennie, in {\it Hadronic Interactions of Electrons and Photons}, edited by J.~Cummings and H.~Osborn (Academic, New York/London, 1971), p. 321.

\bibitem{SIS} G.~Wolf, W.~Cassing and U.~Mosel, Nucl. Phys. A {\bf 552}, 549 (1993); S.~Teis, W.~Cassing, M.~Effenberger, A.~Hombach, U.~Mosel and G.~Wolf, Z. Phys A {\bf 356}, 421 (1997); A.~Hombach, W.~Cassing, S.~Teis and U.~Mosel, Eur. Phys. J. A {\bf 5}, 157 (1999).

\bibitem{Eff99a} M.~Effenberger, E.~L.~Bratkovskaya, W.~Cassing and U.~Mosel, Phys. Rev. C {\bf 60}, 027601 (1999).

\bibitem{res-reg} M.~Effenberger, A.~Hombach, S.~Teis and U.~Mosel, Nucl. Phys. A {\bf 614}, 501 (1997); J.~Lehr, M.~Effenberger and U.~Mosel, Nucl. Phys. A {\bf 671}, 503 (2000).

\bibitem{Eff00} M.~Effenberger and U.~Mosel, Phys. Rev. C, {\bf 62}, 014605 (2000).

\bibitem{Bia96Muc99} N.~Bianchi {\it et al.}, Phys. Rev. C {\bf 54}, 1688 (1996); V.~Muccifora {\it et al.}, Phys. Rev. C {\bf 60}, 064616 (1999).

\bibitem{Fal00} T.~Falter, S.~Leupold and U.~Mosel, Phys. Rev. C {\bf 62}, 031602 (2000).

\bibitem{Fal01} T.~Falter, S.~Leupold and U.~Mosel, Phys. Rev. C {\bf 64}, 024608 (2001).

\bibitem{Eff99b} M.~Effenberger, E.~L.~Bratkovskaya and U.~Mosel, Phys. Rev. C {\bf 60}, 044614 (1999).

\bibitem{And93} B.~Anderson, G.~Gustafson and Hong~Pi, Z. Phys. C {\bf 57}, 485 (1993).

\bibitem{Bau78} T.~H.~Bauer, F.~Pipkin, R.~Spital and D.~R.~Yennie, Rev. Mod. Phys. {\bf 50}, 261 (1978).

\bibitem{Ack00} K.~Ackerstaff {\it et al.}, Phys. Rev. Lett. {\bf 82}, 3025 (1999).

\bibitem{Glauber-models} J.~H\"ufner, B.~Kopeliovich and J.~Nemchik, Phys. Lett. B {\bf 383}, 362 (1996); J.~H\"ufner and B.~Kopeliovich, Phys. Lett. B {\bf 403}, 128 (1997); R.~Engel, J.~Ranft and S.~Roesler, Phys. Rev. D {\bf 55}, 6957 (1997); G.~Kerley and G.~Shaw, Phys. Rev. D {\bf 56}, 7291 (1997); A.~Pautz and G.~Shaw, Phys. Rev. C {\bf 57} 2648 (1998); T.~Renk, G.~Piller and W.~Weise, Nucl. Phys. A {\bf 689}, 869 (2001); B.~Kopeliovich, J.~Nemchik, A.~Sch\"afer and A.~V.~Tarasov, Phys. Rev. C {\bf 65}, 035201 (2002).

\end{thebibliography}
\end{document}